\documentclass[prl,floatfix,twocolumn]{revtex4}
\usepackage{epsf}

\newcommand{\be}{\begin{equation}}
\newcommand{\ee}{\end{equation}}

\begin{document}

\title{Does the crossover from perturbative to nonperturbative physics
in QCD become a phase transition at infinite N ?}
\author{J. Kiskis}
\affiliation{
Department of Physics, University of California, Davis, CA 95616
\\{\tt kiskis@physics.ucdavis.edu}}
\author{R. Narayanan}
\affiliation{
Department of Physics, Florida International University, Miami,
FL 33199\\{\tt rajamani.narayanan@fiu.edu}}
\author{ H. Neuberger}
\affiliation{
Rutgers University, Department of Physics
and Astronomy,
Piscataway, NJ 08855\\{\tt neuberg@physics.rutgers.edu}
}

\begin{abstract}
We present numerical evidence that, in the planar limit,
four dimensional Euclidean Yang-Mills theory 
undergoes a phase transition on a finite symmetrical four-torus 
when the length of the sides $l$ decreases to a critical value $l_c$.
For $l>l_c$ continuum reduction holds so that at leading order in
$N$, there are no finite size effects in Wilson and Polyakov 
loops.
This produces the exciting possibility of solving 
numerically for the meson sector
of planar QCD at a cost substantially smaller than 
that of quenched $SU(3)$.

\end{abstract}

\maketitle

{\bf 1. Introduction. }
Nonabelian gauge theories in four dimensions interact strongly at large
distances and weakly at short distances. The major conceptual achievement
of lattice gauge theory has been to show that the continuum limit
contains both regimes and interpolates between them in a smooth manner.
At present, only numerical simulations can bridge these two regimes.

It has been a long held hope \cite{thooft} that the 
task would simplify at infinite
number of colors $N$. In its most optimistic version, the hope is that
somebody will come up with an analytic solution to the $N=\infty$, 
``planar'', limit at all scales, and that it also will be feasible to compute
$\frac{1}{N}$ corrections. We are far from attaining this goal, but
there has been recent progress on similar, albeit vastly more constrained
theories \cite{strings}, the most notable example being 
${\cal N}=4$ SUSY YM. The exact explicit solution of
the planar limit in this case is limited to extreme cases, the simplest 
among them being
when the 't Hooft coupling ($\lambda=g^2_{\rm YM} N$) is taken to infinity.
The general case has been mapped into a 
two dimensional field theoretic
problem which is not yet fully understood. The
$\frac{1}{N}$ corrections will
come from the interactions in IIB string theory expanded 
around an $AdS_5 \times S_5$
background stabilized by a non-trivial RR flux. 
At large $\lambda$ many answers can be obtained by relatively simple
calculations of small perturbations around this supergravity background. 

In pure YM, we do not have a free coupling constant: instead there is 
nontrivial scale dependence reflecting the breaking of
conformal invariance. A simple way to visualize the situation is to
consider Bjorken's ``femto-universe'' \cite{bjorken}, 
where one studies the Hamiltonian of QCD restricted to a
small three dimensional box of side $l$.
Equivalently, one can not only shrink the system (which takes care
of linear momentum) but also raise its temperature (which deals with
frequencies). The latter situation is best described by Euclidean field
theory on a four torus. A remnant of Lorentz invariance is
preserved when the temperature and linear size of the torus
are related so that in Euclidean space the four torus has sides
of equal length. As $l$ is varied, the bulk physics of the system 
is perturbative for $l\Lambda_{\rm QCD} <<1$ and nonperturbative for
$l\Lambda_{\rm QCD} >> 1$. Thus, the role of the coupling is 
taken up by $l$.

In this paper, we consider 
$SU(N)$ pure gauge theory in the planar limit on a four-torus of side $l$.
The most basic question is whether, as a function of $l$, this system
undergoes any phase transitions at $N=\infty$. There can be such
transitions because $N$ {\it equals} infinity: for finite $N$ all 
these transitions will be smoothed out into crossovers. 
It is possible that when the torus is taken to infinite four-volume
in a specific way, the crossovers
become transitions even at finite $N$. How the various transitions
merge into a coherent picture is a question that we have only begun to
explore. 

Recent work in three dimensions \cite{cek3} has 
shown that when $l$ is decreased from
infinity, there will be a transition at $l=l_c$. For $l>l_c$, the
system realizes a continuum version of Eguchi-Kawai 
reduction \cite{ek}, whose
salient feature is that expectation values of arbitrary Wilson loops
are exactly $l$-independent once $N=\infty$. This continuum Eguchi-Kawai
reduction is a property that is natural for a system of free strings
but more difficult to understand in field theoretical terms where,
in order to erase the perturbative $l$-dependence, 
one is required to enlist an averaging over a moduli space of minima
of the classical action. In this paper, we show that essentially the
same effects work in four dimensions as well.

{\bf 2. General phase structure. }
We used a single plaquette Wilson action as the cost of simulation
increases with the number of colors 
$N$ as $N^3$, making the simplicity of the action
a relevant resource consideration. We worked on tori 
whose sides consist of $L$
lattice sites. The physical side length is $l=aL$, where $a$ is the
lattice spacing. $L$ was
varied from $1$ to $10$. The parameter $N$ was chosen to be a prime 
number, taking the values $N=23,31,37$. The preference for prime
$N$ is because the transition we found has to do with $Z(N)$ groups,
and for prime $N$, $Z(N)$ does not have subgroups that could confuse
the picture. In one case, we did a simulation at $N=27$ and found 
nothing unusual, so it is possible that the choice of
prime $N$ values was unnecessary. 

\begin{eqnarray}
S=\frac{\beta}{4N}\sum_{x,\mu\ne\nu} Tr[ U_{\mu,\nu}(x)
+U_{\mu,\nu}^\dagger (x) ] \\
U_{\mu,\nu}(x)=U_\mu (x) U_\nu (x+\mu) U_\mu^\dagger (x+\nu) 
U_\nu^\dagger (x)
\end{eqnarray}
We define $b=\frac{\beta}{N}=\frac{1}{\lambda}$ and take 
the large $N$ limit with $b$ held fixed. As usual, $b$ determines
the lattice spacing $a$. 
The lattice is a symmetric torus
of side $L$. The gauge fields are periodic. $x$ is a four component integer
vector labeling the site, and $\mu$ either labels a direction or denotes
a unit vector in the $\mu$ direction. The link matrices $U_\mu(x)$ 
are in $SU(N)$.

There is a $Z^4(N)$ symmetry under which 
\begin{equation}
U_\mu (x)\rightarrow e^{\frac{2\pi\imath k_\mu }{N}} U_\mu (x)
\end{equation}
for all $x$ with $x_\mu = c_\mu$. The integers $c_\mu$ are 
fixed, and the integers $k_\mu$ label the elements of
the group; $c_\mu , k_\mu =0,1,..,L-1$. Changing the $c_\mu$'s  
amounts to a local gauge transformation.  

Polyakov loops are denoted by $P_\mu (x)$ and defined by:
\begin{equation}
P_\mu (x)=U_\mu (x) U_\mu (x+\mu) U_\mu (x+2\mu)..U_\mu(x+(L-1)\mu)
\end{equation}
Under the above symmetry, $P_\mu (x)$ gets multiplied by a phase.
The gauge invariant content of  $P_\mu (x)$ is its set of 
eigenvalues (the spectrum)
$e^{\imath\theta^P_i},~i=1,2...,N$. The ordering is not gauge invariant,
and there is a constraint that $\det P_\mu (x) =1$. Under a $Z(N)$
transformation, the set of eigenvalues is circularly shifted by a fixed
amount. The spectra of $P_\mu (x)$ and of $P_\mu (x+j\mu)$ are
the same for all $j=0,1,2....,L-1$. Wilson loops are defined similarly to 
Polyakov loops, only they are invariant under the $Z^4(N)$. 
Often we shall speak about the angles $\theta^P_i$, thinking about them
round a circle and referring to them also as the ``spectrum''. 

At a given $L$, we increase $b$ gradually, until a point is reached
where one of the four $Z(N)$ factors, acting 
in a randomly picked direction $\mu$, breaks spontaneously. The breaking
is reflected by a change in the spectra of $P_\mu (x)$ away from a 
form symmetric under circular shifts. This happens when 
a gap larger than $\frac{2\pi}{N}$ opens up in the 
the angle spectrum at some random location
round the circle. This event can be detected in various ways.

At infinite $N$, six phases are encountered 
as $b$ is varied from zero to infinity on a lattice
of size $L^4$ so long as $L \ge 9$. 
In the range $0 < b < 0.36$ 
the system is in a ``hot'' phase (denoted by ``0h''),
where the $Z^4(N)$ is preserved and the $1\times1$ 
Wilson loop has no gap in its spectrum.
As $b$ increases one goes into a ``0c'' 
phase, where the $Z^4(N)$ 
symmetry still is preserved,
but the $1\times 1$ Wilson loop now has a gap in its spectrum. 
The gap is centered at the point $-1$ on the unit circle, so that
charge conjugation is also preserved. 
Next one goes into a
``1c'' phase, where exactly one factor of the global $Z^4(N)$ is broken. 
As $b$ increases further, 
additional factors of $Z^4(N)$ successively break until the phase ``4c'' 
is reached, which extends all the way to
$b=\infty$.  For $5\le L\le 8$ 
the phases ``0c'' and ``1c'', including the transition
between them, can be extended 
downwards in $b$, as metastable phases, into 
the ``0h'' phase region. Thus, using metastability,
we can extend most of the phase structure of interest
from $L\ge 9$ to $L\ge 5$. 
For $L\le 4$ the stable``0c'' phase is 
``squeezed'' out, and we are left with only five phases. 
The case $L=1$ is the original Eguchi-Kawai model. 
All this holds with 
Wilson's single plaquette action and might change 
with a different lattice action.  However, physical properties that survive  
the continuum limit should be insensitive to the choice of action. 

The ``0c'' phase is the most
interesting phase because there planar $QCD$ exhibits confinement 
and stringy behavior at large distances 
and field theoretic asymptotic freedom
at short distances.  
The phase ``4c'' is the phase where planar $QCD$ is well 
described by Bjorken's femto-universe 
heated to high temperature. 

There is little doubt that the ``4c'' phase survives in the 
continuum limit. This means that there
exists a finite range of torus sides $l$ between zero and some 
small scale where planar 
continuum QCD is in a ``4c''  phase. There also is little 
doubt that the ``0h'' phase
does not have a continuum limit, i.e. there
is no finite range of $l$-values in which the continuum system is
in a ``0h'' phase. In other words, the ``0h" phase is a
lattice artifact. 
In this paper, we shall 
present evidence that the ``0c'' phase 
does have a continuum limit, describing the
system in the range $\infty > l >l_c$. 
More work will be needed to complete the continuum
phase diagram and see in detail how 
the system goes from the ``0c'' phase all the way to
the ``4c'' phase as the torus is shrunk in size.  

{\bf 3. Numerical method and results. } We simulated the 
system using the Monte Carlo method 
employing heat bath updates and overrelaxation updates. The heat-bath
updates amounted to sequential $SU(2)$ updates, going over a
set of $\frac{N(N-1)}{2}$ of $SU(2)$ subgroups identified by choosing
two distinct integers between $1$ and $N$ \cite{cabmar}. 
For most of the values of $b$ we used, each $SU(2)$ update was done
using the Kennedy-Pendelton method \cite{kenpen}. For few small
values of $b$ we used the original Creutz method \cite{creutz2}.
The cost of a heat-bath update goes as $N^3$ as $N$ increases.

The overrelaxation update was a full $SU(N)$ 
update \cite{creutz, phillipe1, phillipe2} 
and had a comparable cost.  Our implementation went as follows:

The portion of the action $S$
that depends on a particular link matrix, denoted by $U$, $S_R(U)$, is
given by
\begin{equation}
Tr \left [ U\Sigma \right ] =\frac{1}{2} [ S_R (U)+i S_I (U) ]
\end{equation}
with real $S_{R,I}(U)$. 
$\Sigma$ are the ``staples'', a positive number (the coupling)
times a sum of simple unitary matrices. $\Sigma$ is determined
by $U$ and when this is not evident from the context we shall
use the notation $\Sigma_U$.  With probability one, $\Sigma$ 
has non-zero determinant permitting a unique definition of a unitary
matrix $V_\Sigma$:
\begin{equation}
V_\Sigma=\frac{1}{\sqrt{\Sigma\Sigma^\dagger}}\Sigma
,~\det V_\Sigma=\frac{\det\Sigma}{|\det\Sigma |}\equiv e^{i\Phi_\Sigma}
\end{equation}

$V_\Sigma$ is calculated as follows: $\Sigma\Sigma^\dagger$ and 
$\Sigma^\dagger \Sigma$ are diagonalized using the Householder method.
The eigenvalues are distinct with probability one and make up
a positive diagonal matrix ${\bf D}$. The diagonalizing matrices
provide two representations of ${\bf D}$:
${\bf D}=X\Sigma^\dagger \Sigma X^\dagger = Y\Sigma\Sigma^\dagger Y^\dagger$ 
and, finally, we end up with $V_\Sigma = Y^\dagger X$.

Taking $\Phi_\Sigma$ to obey
$\pi \ge \Phi_\Sigma > -\pi$ 
we define a new $SU(N)$ matrix by
\begin{equation}
V=e^{\frac{2\imath\Phi_\Sigma}{N}}
V_\Sigma^\dagger U^\dagger V_\Sigma^\dagger
\end{equation}

The update starts by ``offering'' the replacement $U\rightarrow V$.
The new action is given by 
\begin{equation}
S_R (V) = \cos\left ( \frac{2\Phi_\Sigma}{N} \right ) S_R (U) + 
\sin \left ( \frac{2\Phi_\Sigma}{N} \right ) S_I (U)
\end{equation}

For large $N$ we expect $S_R(U)$ to be order $N^2$ and
$S_I (U)$ to be order $N$. This implies that the change in
action is order one and hence there is an order one
probability of making a large move in configuration space. 
In the Metropolis step, the {\it a priori} probability for change is unity
for the $U\to V$ transition and zero for anything else.
Applied twice, this transition becomes the identity; therefore
the Metropolis step only depends on $R$, the ratio of the Boltzmann factors.
\begin{equation}
R=e^{\sin \left ( \frac{2\Phi_\Sigma}{N} \right ) S_I (U)-
\left [ 1- \cos\left ( \frac{2\Phi_\Sigma}{N} \right ) \right ] S_R (U)}
\end{equation}
The acceptance probability for the change is taken
as min\{$1,R$\}; this satisfies detailed balance.

We found that the acceptance rate for the overrelaxed
update was over 95\% for all our $N$ and $b$ values. We employed
a mixture of heat bath and overrelaxation steps of equal amounts. 
A more comprehensive independent study of full $SU(N)$ overrelaxation
has been recently presented in \cite{phillipe2}. 

Our Polyakov loops (as well as various Wilson loops we looked at) 
were built out of $\tilde U_\mu(x)$ matrices,
rather than the original link matrices $U_\mu (x)$. The  
$\tilde U_\mu(x)$ matrices are defined in term of the  $U_\mu (x)$
by an iterative ``smearing'' procedure \cite{ape}. One step in the iteration
takes one from a set $U^{(n)}_\mu (x)$ to a set $U^{(n+1)}_\mu (x)$,
by the following equation:
\begin{eqnarray}
X^{(n+1)}_\mu (x)\equiv\alpha U^{(n)}_\mu (x)+\frac{1-\alpha}{6}
\Sigma_{U^{(n)}_\mu (x)}\nonumber\\
U^{(n+1)}_\mu (x)=X^{(n+1)}_\mu (x)
\frac{1}{\sqrt{[X^{(n+1)}_\mu (x)]^\dagger X^{(n+1)}_\mu (x)}}
\end{eqnarray}
We chose $\alpha=0.45$ and iterated $L$-times:
\begin{equation}
\tilde U_\mu(x)=U^{(L)}_\mu (x)
\end{equation}
This has little effect on Polyakov spectra at smaller $b$ values,
but, after the transition the well known ultraviolet renormalization 
\cite{vergeles} 
of Polyakov loops will reduce all traces $Tr P^k_\mu (x,\mu)$,
effectively making the angle spectrum look more uniform and
making the transition harder to discern. This effect is reduced by
the above smearing. In our three dimensional  
work \cite{cek3} we could do without 
smearing, since the ultraviolet divergence is milder. 
The Polyakov loops in terms of the
smeared links have the same symmetry properties as the Polyakov loops
in terms of the original links and therefore provide perfectly adequate
order parameters for $Z(N)$ symmetry breaking.

\begin{figure}
\epsfxsize = 0.45\textwidth
\centerline{\epsfbox{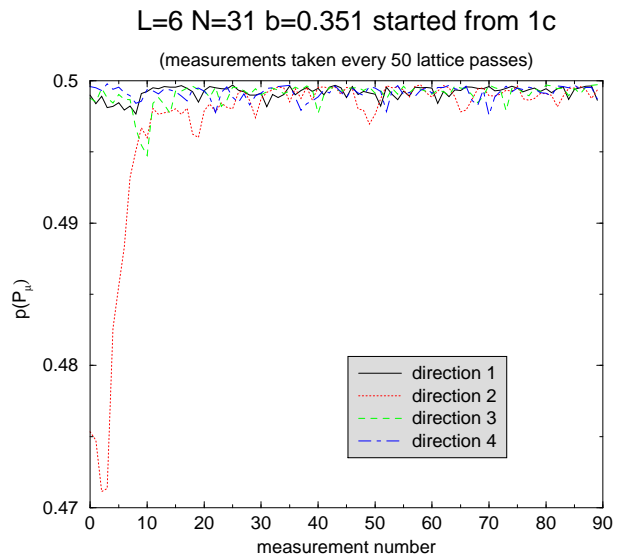}}
\caption{ History of the variable $p(\tilde P_\mu )$ for each direction.
We see the evolution from a state where one of the four $Z(N)$ 
factors is broken to one in which all four are preserved. 
}
\label{c1_to_0c}
\end{figure}

\begin{figure}
\epsfxsize = 0.45\textwidth
\centerline{\epsfbox{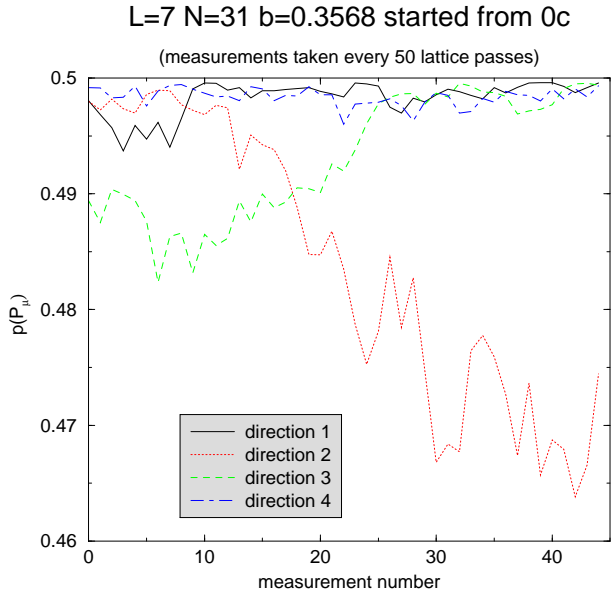}}
\caption{ History of the variable $p(\tilde P_\mu )$ for each direction.
We see the evolution from a state where all four $Z(N)$ factors are preserved
to one where one factor is broken. During the first fifty passes (before the
first measurement) Polyakov loops 
in direction 3 have acquired some structure 
but, ultimately, direction 2 is selected for breakdown and the Polyakov loops
in the other three directions converge to a symmetric state. 
}
\label{c0_to_1c}
\end{figure}

\begin{figure}
\epsfxsize = 0.45\textwidth
\centerline{\epsfbox{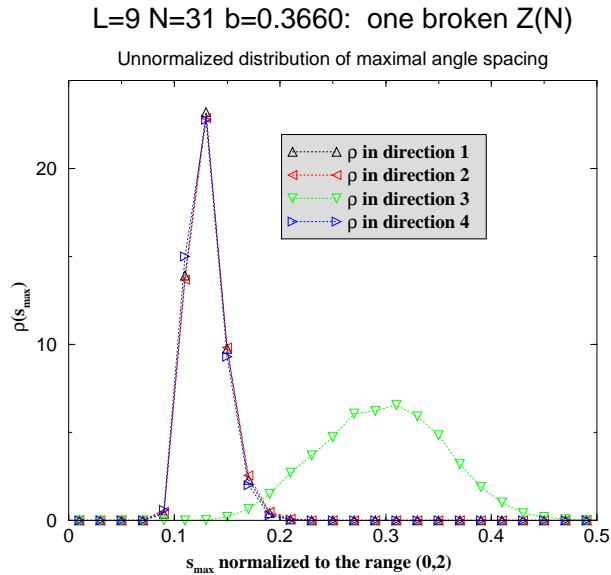}}
\caption{ Here we show the difference between the 
distributions of the largest inter-angle
spacing for smeared Polyakov loops in different 
directions in the phase where exactly
one $Z(N)$ factor is broken. (At other couplings, 
where no $Z(N)$ factor is broken, all 
four distributions look like the three unbroken ones here). 
}
\label{s_dist_1c}
\end{figure}

\begin{figure}
\epsfxsize = 0.45\textwidth
\centerline{\epsfbox{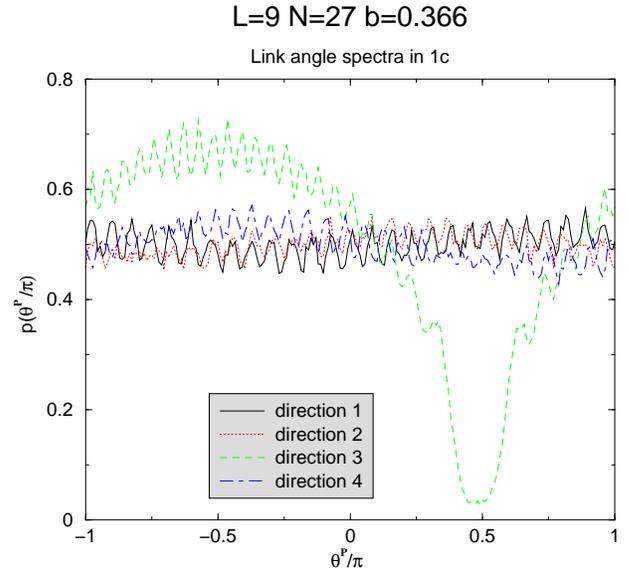}}
\caption{Angle distributions in four directions in the 1c phase. There are
twenty seven periods in the superposed oscillations. The peaks, except  close
to the gap associated with direction 3, are equally spaced. 
}
\label{ex_dist}
\end{figure}

\begin{figure}
\epsfxsize = 0.45\textwidth
\centerline{\epsfbox{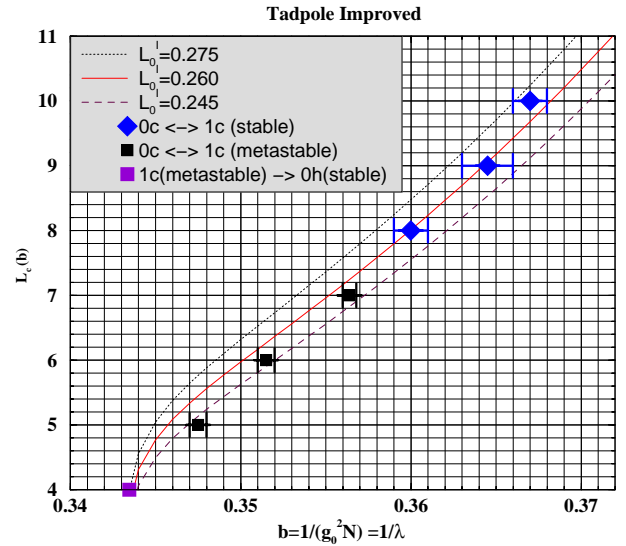}}
\caption{The transition ranges compared to possible two loop
renormalization group curves with tadpole improvement. 
}
\label{scomb}
\end{figure}

\begin{table}
\caption{\label{tab1} Summary of ranges for the ``0c'' to ``1c''
transitions.} 
\begin{ruledtabular}
\begin{tabular}{lcr}
$L$ & $N$ &($b_{\rm min}, b_{\rm max}$)\\
\hline
5 & 31 & (0.3470,0.3480)\\
5 & 41 & (0.3473,0.3485)\\
6 & 31 & (0.3510,0.3520)\\
7 & 31 & (0.3560,0.3568)\\
8 & 23 & (0.3590,0.3610)\\
8 & 31 & (0.3595,0.3605)\\
9 & 23 & (0.3630,0.3655)\\
9 & 27 & (0.3635,0.3650)\\
9 & 31 & (0.3630,0.3660)\\
10& 23 & (0.3662,0.3678)\\
\end{tabular}
\end{ruledtabular}
\end{table}

We looked at several observables. Two were the most useful for identifying
the phase transition. The first is \cite{bhn} 
%\begin{eqnarray}
\begin{equation}
p(\tilde P_\mu) = \frac{1}{N^2} \langle \sum_{i,j=1}^N \sin^2 \frac{1}{2} 
(\theta^{\tilde P}_i - 
\theta^{\tilde P}_j )^2\rangle 
%\nonumber\\
%=\frac{1}{2N^2} \langle [1-|tr {\tilde P}_\mu |^2 ] \rangle 
%\end{eqnarray}
\end{equation}
The averaging is over the 3-plane perpendicular to $\mu$ and 
over configurations. 
Equally spaced
angles respect the $Z(N)$ symmetry in the $\mu$ direction and 
maximize $p$ to 0.5. 
When
the angle-spectrum starts getting modulated and opens a gap, 
$p$ drops below 0.5.
The second observable that we found useful, $\rho(s_{\rm max})$, 
is constructed as follows: 
Among all spacing between adjacent angles round the circle 
select the largest one, $s_{\rm max}$. Its probability distribution,
$\rho (s_{\rm max})$, strongly depends on whether the $Z(N)$ associated with
the direction under consideration is broken or not.  When the $Z(N)$ is not
broken the distribution just reflects universal angle repulsion. However,
when a gap opens in the spectrum, it dominates the distribution. 
Of course, looking directly at the histograms of the angles associated with
individual directions remains the most direct way to 
observe the behavior of the system. Figures \ref{c1_to_0c} and  \ref{c0_to_1c} show 
examples of the evolution of $p$ when the run 
started from a typical configuration in the wrong 
phase. Figure \ref{s_dist_1c} shows an example of the 
difference between the distribution of the 
maximal level spacing in the direction corresponding to a 
broken $Z(N)$ and the maximal level spacing distributions in the
other directions, whose corresponding $Z(N)$'s are unbroken.

Figure \ref{ex_dist} shows an example of the angle 
distributions  associated with the four independent directions when 
the system is in the ``1c'' phase. The superposed oscillations reflect
the global $SU(N)$ constraint on angle values. For a $SU(N)$ matrix
drawn with Haar probability measure a relatively straightforward calculation
gives, with $\xi=\frac{\theta^P}{\pi}$:
\begin{equation}
p(\xi)d\xi=\left [ \frac{1}{2}+\frac{(-1)^{N-1}}{N}\cos(N\pi\xi)\right ] d\xi
\end{equation}
Thus, the expected swing between minima and maxima should be $\frac{2}{27}$
in our plot and this fits more or less. Also, because 27 
is an odd number, we have a maximum, rather than a minimum, at $\xi=0$.

The ``0c'' to ``1c'' transition also breaks hypercubic invariance 
since one direction is randomly selected
by the breaking. The  ``0c'' to ``1c'' is transition 
most likely is of first order,
and therefore generates hysteresis cycles for relatively short runs. 
Our runs were of the order of few
thousands and did not allow a very precise identification 
of the location of the transition or a definitive determination of its
order. We ran hysteresis cycles looking for two extremes which 
determine the range we believe
the true transition is in. At the first extreme we start 
from a configuration typical of the ``1c'' phase, 
and see that the system disorders, 
restoring the remaining $Z(N)$ factor.  This is what happens
in Figure \ref{c1_to_0c}. 
For our larger volumes ($L=9$ and $L=10$) 
we try to find the largest $b$ where the configuration 
evolves in this way within at most
3,000-4,000 passes over the lattice. At the other extreme 
we start from a totally
symmetric configuration and observe the system evolving 
into a ``1c'' phase, like in Figure \ref{c0_to_1c}. 
Here we try to find the smallest $b$ where this scenario is realized. 
The ranges in $b$ so obtained were of length somewhere 
between $1\times 10^{-3}$ and $3\times 10^{-3}$.
A few checks, performed by varying $N$ at fixed $L$ , 
showed that the finite $N$ effects were at most
of order $5\times 10^{-4}$. With this 
accuracy, we were able to check whether the
location of the transition $b_c (L)$ varies with $L$ 
in a way compatible with asymptotic freedom.
Our numerical work used up about one year's worth of
time on a dedicated desktop PC with a modern processor
and 2GB of memory.
With this rough map of the phases in place, one could 
proceed to finer determinations, but this
would require one or two orders of magnitude more computer time. 

If the transition we are searching for truly is a continuum phenomenon, 
the inverse of the function $b_c (L)$, $L_c(b)$, should behave for $b\to\infty$ as:
\begin{equation}
L_c(b)~\sim ~L_0 \left ( \frac {11}{48\pi^2 b} \right )^{\frac{51}{121}} 
e^{\frac{24\pi^2 b}{11}}
\end{equation}
The asymptotic regime is not reached at $L\sim 10$, 
but by the ``tadpole'' \cite{lepage} replacement
\begin{eqnarray}
b\rightarrow b_I \equiv b e(b)~~~~~~~e(b)=\frac{1}{N} 
\langle Tr U_{\mu,\nu} (x) \rangle \\
L_c(b)~\sim ~L^I_0 \left ( \frac {11}{48\pi^2 b_I }\right )^{\frac{51}{121}} 
e^{\frac{24\pi^2 b_I}{11}}
\end{eqnarray}
the asymptotic behavior is supposed to set in much earlier \cite{teperolder}. 
The numerical effect the replacement of $b$ by $b_I$ has is summarized by
the approximate relation $\delta b_I \sim 1.3 \delta b$ 
which holds in the vicinity
of the transition at $L=9$.  With $b$ replaced by $b_I$, the 
theoretical curve becomes 
somewhat steeper.
The plaquette expectation
value, $e(b)$, is taken on the symmetric side of the transition 
and using the MC data,
can be well fitted in the range of interest by:
\begin{equation}
e(b)\approx \frac{1+\frac{a_0}{b}+\frac{a_1}{b^2}}{1+\frac{a_2}{b}+\frac{a_3}{b^2}}
\end{equation}
When $b\to\infty$, $e(b)=1$ at leading order in $\frac{1}{b}$ and $b_I=b$. 
We varied $L$ between $4$ and $10$ and $b$ between $.344$ and $.366$. 
$e(b)$ was reasonably well fit by
$a_0 = -0.58964,~a_1 =  0.08467,~a_2 = -0.50227,~a_3 =  0.05479$.  
Hence,
\begin{equation}
L_0=L_0^I e^{\frac{24\pi^2 (a_0 -a_2 )}{11}}\approx 0.1524 L_0^I
\end{equation}

The ranges we determined for the ``0c'' to ``1c'' transitions 
are reasonably well described by a range of $L^I_0$ 
constants between 0.245 and 0.275. 
Figure \ref{scomb} shows the ranges we established on a plot together with
lines representing the tadpole improved two loop renormalization formula with 
different amplitudes.  The relative 
consistency of this fit constitutes our numerical 
evidence that the transition is physical, 
occurring in the continuum at a finite scale. 
The relevant numbers for the ``0c'' to ``1c'' transition ranges that
went into figure \ref{scomb} are collected in Table \ref{tab1}.

Let us first discuss all the stable phases, 
ignoring the metastable ones.
In addition to the ``0c'' to ``1c'' 
transition we have been 
focusing on, the system also undergoes
a lattice transition from ``0h'' to an ``Xc''. 
For small volumes, X will be large than 0, but, starting with
$L=9$, X$=0$. 
The ``0h'' to ``Xc'' (X$\ne 0$) transitions 
at smaller $L$ values are strongly first order 
and occur at  $b=b_{\rm BULK}$.  
A similar transition also occurs at $L=\infty$ for any $N\ge 5$ 
\cite{creutz3}. 
In this case this bulk transition is well known to be associated 
with a large jump in $e(b)$  at $b_{\rm BULK}$. 
At finite $N$ the transition does not break any symmetry.

At $N=\infty$, and any finite but large enough $L$, the location of the 
bulk transition can be estimated with the
help of \cite{campost} to occur at $b_{\rm BULK} = 0.3600$. 
(For an earlier determination, see \cite{okawa}). 
So long as $L>8$, the bulk transition again breaks
no symmetry: It takes the system from the ``0h'' phase to the ``0c'' phase. 
Again, $e(b)$  undergoes a significant jump at 
$b_{\rm BULK}$.  In addition, with $N=\infty$, the average eigenvalue 
distribution of the plaquette variable
now undergoes also a qualitative change, 
opening a gap at angle $\pi$ when $b$ 
increases through $b_{\rm BULK}$.

We have investigated the $L=1$ case in some detail. This case is
special, and the algorithms we used are somewhat different; since
this is a bit of a side issue, we shall not elaborate in detail.
We found evidence for five stable phases. Examples of firmly determined
couplings in each phase are: $b$=0.150 in ``0h'', 
$b$=0.205 in ``1c'', $b$=0.235 in ``2c'', $b$=0.275 in ``3c'' and
$b$=0.320 in ``4c''. 
We also found approximate locations for the transitions:
the ``0h'' to ``1c'' transition occurs at $b=0.19$, 
the ``1c'' to ``2c'' transition occurs at $b=0.22$, 
the ``2c'' to ``3c'' transition occurs at $b=0.26$ and 
the ``3c'' to ``4c'' transition occurs at $b=0.30$.

It is easy to keep the system in ``cold'' (``Xc'') 
phases even for $b<b_{\rm BULK}$. These 
phases are in principle metastable. 
However, in practice, for large enough $N$, ($N\ge 20$), 
the ``Xc'' phases are very stable in a Monte Carlo simulation. 
This makes it
possible to investigate the ``0c'' to ``1c'' transitions 
of interest also for $L\le 7$. 
All our values for $b_c (L)$ for $L\le 7$ have $b_c (L)<b_{\rm BULK}$.

{\bf 4. Relation to the finite temperature deconfinement transition. }
Suppose we studied a torus of unequal sides, $L_\mu$. The most plausible
assumption is that again one $Z(N)$ will break 
first as $b$ is increased from the
phase where the entire $Z^4 (N)$ is preserved.  Only now, which $Z(N)$ breaks
will no longer be arbitrary, but, rather, the one associated with the 
direction with the shortest
$L_\mu$ is selected to break first. Call this direction $\mu_0$. 
The arguments from our previous
paper \cite{cek3} say that for $b$'s smaller than this transition point 
there is no dependence
on the parameters $L_\mu$. Hence, up to the transition, all $L_\mu$'s 
can be considered
as infinite. 
But, we could equally well think about the $L_\mu$ with 
$\mu\ne\mu_0$ as
infinite, while keeping in mind that $L_{\mu_0}$ is finite. 
This puts the system in a situation considered in \cite{gandn}. 
Then, there would be a transition as
$b$ is increased, even for finite $N$. This would be a finite 
temperature transition, which
is first order at large $N$, and has a finite limit at $N=\infty$, 
$T_c$ \cite{teperolder}. The
simplest consistent assumption is that $l_c=\frac{1}{T_c}$.
This is in full agreement with the viewpoint of the authors of 
\cite{teperrecent}. 

In reference \cite{teperolder} the string tension in 
lattice units is found to behave
approximately as follows:
\begin{equation}
\frac{1}{\sqrt{\sigma}} \sim \frac{1}{\sqrt{\sigma^I_0}} 
\left ( \frac {11}{48\pi^2 b_I} \right )^{\frac{51}{121}} 
e^{\frac{24\pi^2 b_I}{11}}
\end{equation}
With a simple minded extrapolation to $N=\infty$
we obtain from  \cite{teperolder} $ \sqrt { \sigma^I_0 }  
\approx  6.05$. Combining this with 
our value of $L^I_0\approx 0.26$ we 
find $\frac{1}{L\sqrt{\sigma}}\approx 0.64$.
The most up to date value for the infinite $N$ value 
of $\frac{T_c}{\sqrt{\sigma}}$ can
be found in \cite{teperrecent}. It is about $0.60$. Thus, 
$l_c=\frac{1}{T_c}$ is consistent with what we know to 
date, but the evidence is not
compelling. 

The special physical effects surrounding the finite 
temperature transition in pure YM
in the planar limit were first discussed in \cite{thorn}.
More aspects have been studied in \cite{morefinitetemp}.
Earlier numerical studies of the infinite $N$ finite temperature transition
in $SU(N)$ gauge theories can be found in \cite{morenumfinitetemp}.

{\bf 5. Preserving $l$ independence in the meson sector. }
We have emphasized the volume independence of the pure gauge theory
in the large $l$ phase. It is natural to ask whether fermions moving in the
gauge backgrounds typical to this phase also will behave as
if the volume were infinite. This question needs to be sharpened because we
are considering here only a finite number of flavors, which makes the fermions
``quenched'' as a result of the large color ($N$) limit. The fermions simply provide
definitions for particular nonlocal gauge invariant 
observables, but do not influence 
the distribution of the gauge background. The $l$ independence holds 
only for single traces of Wilson loops,
and there is a way even for a very large loop to fold up 
into the $l^4$ torus. But a trace
of the product of a fermion by an antifermion propagator 
only depends on the end points,
not on a path, so there seems to be no way to describe a separation 
that would not fit into the torus.

However, there is a trick which seems to allow the definition 
of fermionic observables
on the finite torus which nevertheless describe propagation at larger distances. 
This trick is at the heart of our proposed shortcut to the 
planar limit in the meson
sector. It is a simple generalization of work in \cite{gk, bhn}. This 
produces a prescription 
for calculating  $q\bar q$
properties in the large $N$ limit while preserving volume independence. 
At the diagrammatic
level it is easy to understand what we have done \cite{gk}, but, to be sure, 
the arguments supporting this construction are far from rigorous.  
For this reason, we have
undertaken an extensive test in two dimensions \cite{2d}, which came out favorable.
We believe that this provides sufficient grounds to go ahead and see what 
happens in four
dimensions. 

To be concrete, let us consider the scalars 
$M(x)=\frac{1}{\sqrt{N}}\bar\psi\chi (x)$ and 
$\bar M(x)=\frac{1}{\sqrt{N}}\bar\chi\psi (x)$. 
These meson fields are color singlets. 
$x$ and $y$ are sites on an $L^4$ lattice. 
Normally we would expect only momenta $K_\mu=\frac{2\pi}{L} k_\mu$ with 
$k_\mu=0,1,...,L-1$ to be accessible. We claim that 
large $N$ reduction makes it possible
to interpret data obtained on the $L^4$ lattice as 
providing predictions for momenta
on an $(NL)^4$ lattice, at leading order in the 
$\frac{1}{N}$ expansion. The momenta are
now written as $K+Q$ where $K$ is as before, 
and $Q_\mu=\frac{2\pi}{NL} q_\mu$ with
$q_\mu=0,1,...,N-1$. We are after an expression 
for the meson-meson propagator,
$S(K+Q)$. We first define a shifted link field $U^{(q)}_\mu (x)$ by 
\begin{equation}
U^{(q)}_\mu (x) = e^{\frac{2\pi\imath}{NL} q_\mu} U_\mu (x)
\end{equation}
and denote the fermion $\psi - \bar\psi$ and $\chi - \bar\chi$
propagators on the lattice, in a given gauge background, $\{U\}$, by
$G(x,y,\{U\})$. The shifted gauge field links are not in $SU(N)$.
Let us consider the collection of all Wilson and Polyakov loops
made out of the $U^{(q)}_\mu (x)$ variables as elementary links.
Before taking the trace all the unitary matrices are back in
$SU(N)$ and could have been obtained from elementary links that
also are all in $SU(N)$. This works because
our observables are gauge invariant even under $U(1)$ gauge
transformations that take the links out of $SU(N)$. 
The dependence on the integers $q_\mu$ 
is removable by a $Z^4(N)$ symmetry transformation. Thus, so long
we treat a fermionic observables that depends on a single gauge field
background there is no dependence on $q_\mu$ so long the $Z^4(N)$ symmetry
is not broken. In other words, we have to be in the ``0c'' phase. 
The $q_\mu$ asume their
role as momentum ``gap fillers'' only when we consider an observable
that depends on different gauge field backgrounds. 
The Dirac and color indices of $G(x,y,\{U\})$ are suppressed. 
We now define the quantity
\begin{equation}
R(x,y;\{ U\}, q ) = -\frac{1}{N} Tr [ G(x,y,\{U^{(q)}\}) G(y,x,\{U^{(0)}\}) ]
\end{equation}
The trace in the above equation sums over color and Dirac indices. 
The key formula for $S(K+Q)$ is given by:
\begin{equation}
S(K+Q)=\sum_x e^{\frac{2\pi\imath}{L} k(x-y)} \langle R(x,y;\{ U\}, q )
\rangle_{\{U\}}
\end{equation}
The averaging over $\{U\}$ restores translational invariance. 

If we are not in the ``0c'' phase the procedure fails. One could try
to impose a ``0c'' phase by quenching the links on a $1^4$ lattice, but 
implementing the additional averaging puts too big a burden on the numerics and
is likely less practicable than the procedure we propose here. 

{\bf 6. Conclusions. }
Our previous work \cite{cek3} and the present paper make a plausible case for
the following scenario: 
Planar QCD on a torus of side $l$ has a nontrivial phase 
structure as a function of $l$.
When $l$ decreases from $\infty$ to any $l<l_c$, the system undergoes
a phase transition where the global $Z^4(N)$ symmetry breaks spontaneously.
It is likely that 
$l_c=\frac{1}{T_c}$ where $T_c$ is the infinite $N$ limit 
of the finite temperature deconfinement
transitions of  $SU(N)$ YM theory at finite $N$'s. 
The precise determination of $l_c$, including error estimates, 
and of the existence and locations of other continuum transitions 
needs substantially more numerical effort 
than invested to date,
but is a feasible numerical project. 

A distinctive property of the $l>l_c$ phase is the 
$l$-independence of arbitrary Wilson loops, which provides a 
continuum realization of lattice Eguchi-Kawai reduction. 
While the free energy does not depend on $l$ at leading order in 
a generic, free, string theory with toroidal target space,  
among field theories, only certain gauge theories in the planar limit
can exhibit such a property. 
This scenario, in turn,  leads to the conjecture that 
at $l>l_c$ the $l$ dependence of meson propagators can also be made to 
disappear at leading order in the $\frac{1}{N}$ expansion by quenching, 
providing the opportunity for a numerical shortcut to
the infinite $N$ meson sector of $SU(N)$ YM theory in infinite space-time.

{\bf 7. Acknowledgments.}
R. N. acknowledges partial support by the NSF under
grant number PHY-0300065 and also partial support from Jefferson 
Lab. The Thomas Jefferson National Accelerator Facility
(Jefferson Lab) is operated by the Southeastern Universities Research
Association (SURA) under DOE contract DE-AC05-84ER40150.
H. N. acknowledges partial support at the Institute for Advanced Study
in Princeton from a grant in aid by the Monell Foundation, where most
of this work was done, as well as partial support 
by the DOE under grant number 
DE-FG02-01ER41165 at Rutgers University. 
Scientific Computing facilities at Boston University were used
for part of the numerical computations.

%{\bf 8. References}

\end{document}